\documentclass[12pt,fleqn]{article}
\usepackage{graphicx}
\usepackage{exscale}
\usepackage[intlimits]{amsmath}
\usepackage{amsfonts}
\usepackage{amssymb,amscd}
\usepackage{epsfig}

\newcommand{\beqa}{\begin{eqnarray}}
\newcommand{\eeqa}{\end{eqnarray}}
\newcommand{\beq}{\begin{equation}}
\newcommand{\eeq}{\end{equation}}
\newcommand{\pslash}{p\hspace{-.5em}/\hspace{.15em}}

\begin{document}

\begin{center}
{\LARGE
{\bf Dynamically induced scalar quark confinement}}
\vspace{15mm}

{\large{Reinhard Alkofer$^{1}$, Christian S. Fischer$^{2}$ and\\ Felipe J. Llanes-Estrada$^{3}$}}

\vspace{7mm}

{$^{1}$ Institut f\"ur Physik, Karl-Franzens-Universit\"at,\\ 
Universit\"atsplatz 5, A-8010 Graz, Austria}\\
{$^{2}$ Institut f\"ur Kernphysik, TU Darmstadt,\\ Schlossgartenstrasse 9, 64289 Darmstadt,Germany}\\
{$^{3}$ Dept. F\'{\i}sica Teorica I, Univ. Complutense, Madrid 28040, Spain}
\end{center}
\vspace{2cm}

\centerline{\bf Abstract}
\vspace{3mm}
\small
We employ a functional approach to investigate 
the confinement problem in quenched Landau gauge QCD. We demonstrate 
analytically that a linear rising potential between massive quarks is 
generated by infrared singularities in the dressed quark-gluon vertex. 
The selfconsistent mechanism that generates these singularities is driven 
by the scalar Dirac amplitudes of the full vertex and the quark propagator. 
These can only be present when chiral symmetry is broken. We have thus 
uncovered a novel mechanism that directly links chiral symmetry 
breaking with confinement.
\normalsize

\newpage
\parskip=2mm
\baselineskip=5.5mm

More than thirty years after the identification of non-Abelian gauge theory 
as the appropriate framework to describe the strong interactions
we still lack a satisfactory understanding of the confinement phenomenon.
Isolated particles with nonvanishing colour charges have not been observed in 
nature. This fact is supposed to be encoded in the infrared structure of QCD. 
In the quenched theory a gauge invariant signature of confinement is the area 
law of the Wilson loop at large distances. This behaviour corresponds 
to a linear rising potential between static colour charges  in the 
fundamental representation of the gauge group. This signature has been 
unambiguously verified in lattice gauge QCD, see {\it e.g.\/} ref.\
\cite{Bali:1992ab}. However, the dynamical  mechanism that 
generates the Wilson potential is still elusive. Since the underlying 
long range interaction is provided by gauge dependent objects, this mechanism
may have a different appearance in different gauges. In this letter we present 
such a mechanism for covariant Landau gauge QCD.

The idea of 'infrared slavery', {\it i.e.\/} the notion that   infrared
singularities generate confinement, dates back to  the early seventies
\cite{Weinberg:1973un,Gross:1973ju}. Speculations about the infrared 
behaviour of the
running coupling at that time distinguish two cases:  (A) the coupling develops
an infrared fixed point and (B) the coupling diverges  at small or zero
momenta. The last possibility has been considered as the driving  mechanism for
the generation of infrared singularities and quark confinement. Today there is
ample evidence from functional approaches  that the first possibility is
realised in the Yang-Mills sector of Landau gauge  QCD: the running coupling
freezes out at small momenta 
\cite{vonSmekal:1997is,Lerche:2002ep,Pawlowski:2003hq,Alkofer:2004it}. However,
this does not imply that the Green's functions of Yang-Mills theory are finite
in the infrared: Instead, depending on the number of ghost and gluon legs, many
of the one-particle irreducible Green's functions are indeed singular in the
limit that all
external momenta vanish \cite{Alkofer:2004it}. In this letter we 
demonstrate that these singularities also induce a corresponding singularity
in  the dressed quark-gluon vertex. As a consequence the running
coupling defined from this vertex diverges at vanishing renormalization scale, 
and a linear potential between massive quarks is generated thereby realizing
infrared slavery via an infrared singular quark-gluon vertex. 

Besides confinement the other fundamental property of infrared QCD is 
dynamical chiral symmetry breaking, {\it i.e.\/} the nonperturbative generation
of quark masses from dynamical interaction with gluons. Lattice QCD tells us
that the chiral phase transition and the  deconfinement transition take place
at a similar temperature \cite{Karsch:2003jg}.  This suggests a relation between
confinement and dynamical chiral symmetry breaking, which, however, has not
been clarified in detail yet. In this letter  we present, based mostly on an
analytical calculation, an explicit  mechanism that links these two phenomena. 

The functional approach we employ for our investigation is given by the tower
of Dyson-Schwinger  equations for the one-particle irreducible (1PI) Green's
functions of QCD \cite{Roberts:1994dr,Alkofer:2000wg,Fischer:2006ub}.  
A great advantage of this continuum
based formulation as compared to lattice QCD is the analytical access to the
infrared behaviour of these functions without finite-volume effects necessarily
present in lattice calculations. As has been demonstrated for the propagators
of three-dimensional Yang-Mills theory extremely large lattices are mandatory
to  access the small momentum region where the infrared asymptotic scaling can
be  identified \cite{Cucchieri:2003di}. The computational costs for a
corresponding investigation in four dimensions exceeds current possibilities. 
This problem is even more severe for higher $n$-point functions. This
underlines the necessity for a  continuum based approach to the Green's
functions of QCD.

To begin with we shortly summarise previous results for the infrared behaviour
of the Green's function in Landau gauge Yang-Mills theory. Gauge fixing in this
framework is performed by the standard Faddeev-Popov method supplemented by
auxiliary conditions such that the generating functional consists of an
integral over gauge field configurations that are contained in the first Gribov
region. (The feasibility of this method has been justified in a framework
employing stochastic quantisation  \cite{Zwanziger:2003cf}.) The resulting
Dyson-Schwinger equations for  1PI-Green's functions have been solved
analytically in the infrared to all orders in a skeleton expansion ({\it
i.e.\/} a loop expansion using full propagators and vertices)
\cite{Alkofer:2004it}. 
Here we are interested in the case where all external momenta go to zero.
Choosing all momenta proportional to each other and requiring for the largest 
one $p^2 \ll \Lambda^2_{\tt QCD}$ a 
self-consistent solution of the whole (untruncated) tower of DSEs is given by
\cite{Alkofer:2004it} 
\beq \Gamma^{n,m}(p^2) \sim (p^2)^{(n-m)\kappa},
\label{IRsolution} 
\eeq 
where $\Gamma^{n,m}$ stands for the infrared leading
dressing function of a Green's functions with $2n$ external ghost legs and $m$
external gluon legs. The exponent $\kappa$ is known to be positive
\cite{Watson:2001yv}.  An important property of the infrared solution
(\ref{IRsolution}) is the fact that it is generated by exactly those parts of
the DSEs that involve ghost loops \cite{Alkofer:2004it}. In other words: the
Faddeev-Popov determinant dominates the infrared behaviour of  non-Abelian
Yang-Mills theories. Thus an infrared asymptotic theory can be obtained by
`quenching` the Yang-Mills action, {\it i.e.\/} setting $\exp(-S_{YM}) \rightarrow 1$ in the
generating functional \cite{Zwanziger:2003cf}. The solution of this asymptotic
theory is given  by the power laws (\ref{IRsolution}). Interestingly, this
limit is a continuum  analogue of the strong coupling limit of lattice gauge
theory.

Examples of the general solution (\ref{IRsolution}) are given by
the inverse ghost and gluon dressing functions, $\Gamma^{1,0}(p^2) =
G^{-1}(p^2)$ and  $\Gamma^{0,2}(p^2) = Z^{-1}(p^2)$, respectively. 
They are related to the ghost and gluon propagators via
\beq
D^G(p^2) = -\frac{G(p^2)}{p^2} \, , \ \
D_{\mu \nu}(p^2)  = \left(\delta_{\mu \nu} -\frac{p_\mu 
p_\nu}{p^2}\right) \frac{Z(p^2)}{p^2} \, .
\eeq
The corresponding power laws in the infrared are
\beq
G(p^2) \sim (p^2)^{-\kappa}, \hspace*{1cm} Z(p^2) \sim (p^2)^{2\kappa}\,.
\label{kappa}
\eeq
Since $\kappa$ is positive one obtains an infrared diverging ghost propagator,
a behaviour which is necessary to ensure a well defined, {\it i.e.\/} unbroken,
global colour charge \cite{Kugo:1995km}. In Landau gauge an explicit value for 
$\kappa$ can be derived from the observation that the dressed ghost-gluon vertex
becomes (almost) bare in the infrared 
\cite{Taylor:1971ff,Cucchieri:2004sq,Schleifenbaum:2004id}.
One then obtains $\kappa = (93 - \sqrt{1201})/98 \approx 0.595$ 
\cite{Lerche:2002ep,Pawlowski:2003hq,Zwanziger:2001kw}, which implies 
that the 
gluon propagator vanishes in the infrared. A direct consequence of this behaviour 
are positivity violations in the gluon propagator and therefore the confinement 
of transverse gluons \cite{vonSmekal:1997is,Alkofer:2003jj}.

A further important consequence of the solution (\ref{IRsolution}) is the 
qualitative universality of the running coupling in the infrared.  
Renormalisation group invariant couplings can be defined from either of the 
primitively divergent vertices of Yang-Mills-theory, {\it i.e.\/} from the 
ghost-gluon vertex ($gh-gl$), the three-gluon vertex ($3g$) or the four-gluon 
vertex ($4g$) via
\beqa
\hspace*{-2mm}
\alpha^{gh-gl}(p^2) &=& \frac{g^2}{4 \pi} \, G^2(p^2) \, Z(p^2) 
     \hspace*{9mm} \stackrel{p^2 \rightarrow 0}{\sim} \hspace*{2mm} 
     {\bf c_1}/N_c \,, \nonumber\\
\alpha^{3g}(p^2) &=& \frac{g^2}{4 \pi} \, [\Gamma^{0,3}(p^2)]^2 \, Z^3(p^2) 
    \hspace*{2mm} \stackrel{p^2 \rightarrow 0}{\sim}
     \hspace*{2mm} {\bf c_2}/N_c \,,\nonumber\\
\alpha^{4g}(p^2) &=& \frac{g^2}{4 \pi} \, [\Gamma^{0,4}(p^2)]^2 \, Z^4(p^2) 
    \hspace*{2mm} 
    \stackrel{p^2 \rightarrow 0}{\sim} \hspace*{2mm} {\bf c_3}/N_c \,.
     \label{alpha}
\eeqa
Employing the DSE-solution (\ref{IRsolution}) it is easy to see that all three 
couplings approach a fixed point in the infrared. However, the explicit value of
the fixed point, $c_{1,2,3}/N_c$, may be different for each coupling. For a bare 
ghost-gluon vertex one obtains $\alpha^{gh-gl}(0) \approx 8.92/N_c$ \cite{Lerche:2002ep}; 
the other couplings have not been determined yet. Equations (\ref{alpha}) underline
that indeed the fixed point scenario (A) is realised in the Yang-Mills sector of QCD. 
In turn this explains the existence of the power law solutions (\ref{IRsolution}): 
the theory becomes approximately conformal in the far infrared.

We now present a significant extension of this analysis to the quark sector of
quenched QCD. To this end we first choose the masses of the valence quarks to  be
large, {\it i.e.\/}  $m \ge \Lambda_{\tt QCD}$. The remaining scales below
$\Lambda_{\tt QCD}$ are those of the external momenta of the Green's functions.
Here we are primarily interested in the case where all external momenta go
to zero. Then, without loss of generality, one can choose all external
momenta to be proportional to one scale $p^2$, which is small compared to
all other scales in the theory, i.e. $p^2 \ll \Lambda_{\tt QCD}^2$.
One can then employ Dyson-Schwinger equations to search for selfconsistent 
solutions in terms of powers of $p^2$. 
\footnote{Integrals with an infrared power-law enhancement are 
themselves dominated by infrared integration momenta. We illustrate 
this with the following simple example, where $\alpha>1$ is a real 
number. Consider
$$
I(p)=\int_0^\Lambda \frac{dq}{(q+p)^\alpha}= \frac{1}{1-\alpha}
\left( (\Lambda+p)^{1-\alpha} - p^{1-\alpha} \right) \ .
$$
Then, asymptotically for small $p$, 
$$
I(p) \to  \frac{1}{\alpha - 1} \frac{1}{p^{\alpha-1}}
$$
and it is the lower integration limit that contributes. Then we just 
need to subdivide the integration interval 
$$
\int_0^\Lambda = \int_0^{\lambda_1} + \int_{\lambda_1}^{\lambda_2} + 
\dots  \int_{\lambda_{n-1}}^{\lambda_n}
$$
and it is always the lower ($0$) limit of the first subinterval that 
generates the power-law in $p$. Therefore, for low-$p$, the $q$ 
integration is infrared dominated. \\
Another way of putting it is that the integral is divergent with $p=0$, 
and keeping finite $p$ regulates it. Then the integral is dominated by 
momentum scales of order the regulator. 
}

\begin{figure}[t]
\centerline{\epsfig{file=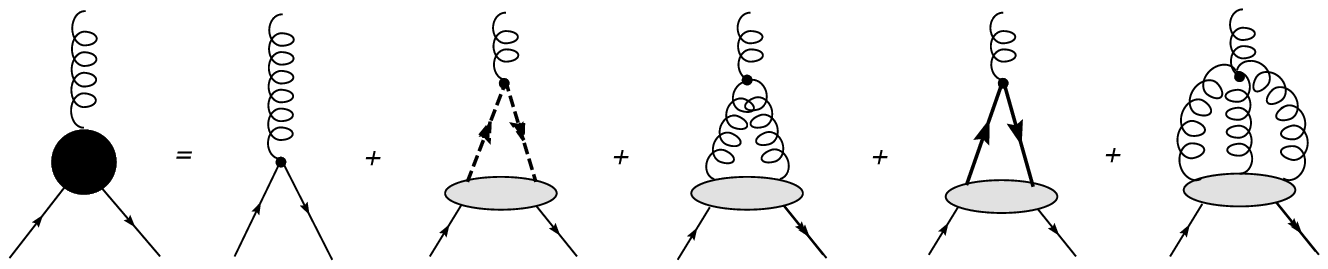,width=80mm}}
\centerline{\epsfig{file=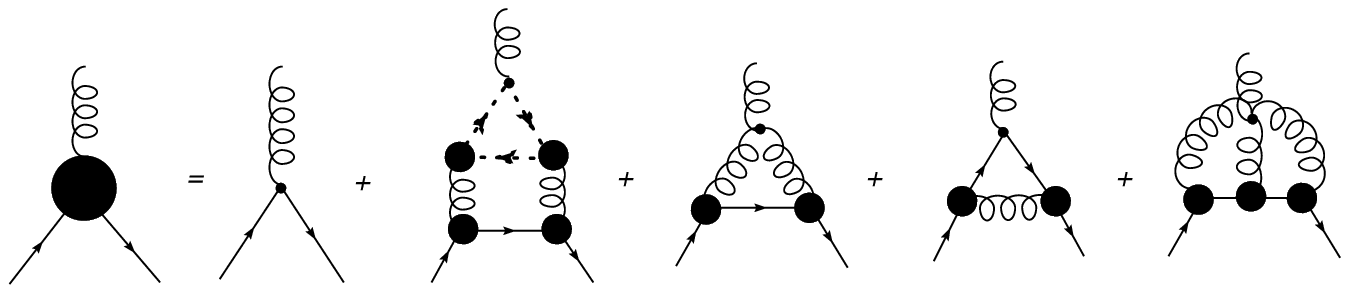,width=80mm}}
\caption{Dyson-Schwinger equation for the quark-gluon vertex. In the first line 
we show the full equation \cite{Marciano:1978su}. The second line shows the 
lowest order of a skeleton expansion, that already reveals the correct power 
counting in the infrared. All internal lines in the diagrams represent fully 
dressed propagators.}
\label{vertex-DSE}
\end{figure}

We apply this method to the Dyson-Schwinger equation for the full  quark-gluon
vertex, given diagrammatically in the first line of Fig.~\ref{vertex-DSE}. In
the second line the higher $n$-point functions of this equation have been
expanded to lowest order in a skeleton expansion in terms of full propagators
and vertices. The dressed quark-gluon vertex $\Gamma_\mu$ can be decomposed in
a basis of twelve tensor  structures, which are given explicitly {\it e.g.\/} in
\cite{Ball:1980ay}. Due to momentum  conservation the vertex depends only on
the two external quark momenta ($p_\mu$, $q_\mu$)  or three Lorentz invariants
($p^2, q^2, p\cdot q$). To analyse the infrared  limit of this vertex in the
presence of only one external scale $p^2 \ll \Lambda_{\tt QCD}^2$, we can set
$q_\mu = 2p_\mu$ without loss of generality.  This leaves us with only four 
possible tensor structures which can be denoted by
\beq
\Gamma_\mu (p) = i g \sum_{i=1}^4 \xi_i (p^2) G_{\mu}^i \label{vertex}
\eeq
with $\xi_i $ being Lorentz and Dirac scalar functions and
\beq
G_{\mu}^1 = \gamma_\mu \,, \ \
G_{\mu}^2 = \hat{p}_\mu \,, \ \
G_{\mu}^3 = \hat{\pslash} \hat{p}_\mu \,, \ \
G_{\mu}^4 = \hat{\pslash} \gamma_\mu\,, 
\eeq
where we have normalised the momentum,
$\hat{p}_\mu =p_\mu / \sqrt{p^2}$, to ease power counting. 
Note that $G_{\mu}^1$ and $G_{\mu}^3$ have an odd number of 
$\gamma$-matrices, whereas $G_{\mu}^2$ and $G_{\mu}^4$ have an even number. 
Therefore $\xi_{2,4}\not=0$ only if chiral symmetry is broken.

The internal loops of the skeleton expansion in Fig.~\ref{vertex-DSE} are 
dominated by loop momenta similar to the external scale $p^2$ due to the 
denominators of the loop propagators. Thus for $p^2 \ll \Lambda_{\tt QCD}^2$ 
the dressing functions of internal propagators and vertices can be approximated 
by the power laws given in (\ref{IRsolution}). We therefore use (\ref{kappa}) 
for the scaling of the ghost and gluon propagators. The ghost-gluon vertex 
scales as a constant in the infrared 
\cite{Taylor:1971ff,Cucchieri:2004sq,Schleifenbaum:2004id}. The internal 
quark propagator lines can be written as
\beq
S(p) = \frac{i \pslash + M(p^2)}{p^2 + M^2(p^2)}Z_f(p^2) \rightarrow 
\frac{i \pslash Z_f}{M^2}+\frac{Z_f}{M}\,. \label{quark}
\eeq
for momenta $p^2 \ll \Lambda_{\tt QCD}^2$, $M=M(p^2 \rightarrow 0) 
\stackrel{\scriptstyle >}{\scriptstyle \sim} \Lambda_{\tt QCD}$ and 
$Z_f=Z_f(p^2 \rightarrow 0)$. Hereby we assume that neither the quark 
mass function $M(p^2)$ nor the wave function renormalisation $Z_f(p^2)$ 
is singular in the infrared \cite{Fischer:2003rp}. This assumption will 
be justified below by selfconsistency arguments. From the two tensor 
structures in (\ref{quark}) the scalar piece turns out to be the leading 
one in the infrared. 

For the internal quark-gluon vertices we can employ the expression
(\ref{vertex}) with any internal momentum as argument: it contains all 
possible types of Dirac structures (vector, scalar, and tensor) and any 
more complicated dependence on external and internal momenta will generate 
the same powers of external momenta in $\xi_{1..4}$ after integration 
(for dimensional reasons). To determine the infrared exponents 
of the quark-gluon vertex we employ the ansatz
\beq
\xi_1 \sim (p^2)^{\beta_1} \,, \ \
\xi_2 \sim (p^2)^{\beta_2} \,, \ \
\xi_3 \sim (p^2)^{\beta_3} \,, \ \
\xi_4 \sim (p^2)^{\beta_4}		\label{vertex-exp}
\eeq
for the scaling of the dressing functions of the quark-gluon vertex with momentum. 
We then substitute this ansatz into the skeleton expansion of the vertex-DSE and 
determine the exponents $\beta_{1..4}$ selfconsistently by matching the left
and right hand sides. The resulting infrared solution is given by 
\beqa
\beta_2 = -1/2-\kappa \,,  && \beta_1,\beta_3 \in [(-1/2-\kappa), (-\kappa)] \nonumber \\
&& \beta_4 \in [(-1/2-\kappa), (1/2-\kappa)]\,. \label{power}
\eeqa
We have checked this analytic solution also in numerical calculations and found that
the case 
\beq
\beta_{1 \dots 4}=-1/2-\kappa \label{power2}
\eeq
is realised. The details of this calculation will be presented elsewhere 
\cite{bigpaper}. 

There are two important remarks here: Firstly, similar to the DSEs in the Yang-Mills 
sector it is the diagram containing the ghost loop that dominates the right hand 
side of the equation in Fig.~\ref{vertex-DSE}. Thus 
also the infrared behaviour of the quark sector is dominated by effects generated
from the Faddeev-Popov determinant.
Secondly, the driving tensor structures of the solution (\ref{power}), (\ref{power2})
are the scalar Dirac amplitude $G^2_\mu$ of the quark-gluon vertex and the scalar 
($Z_f/M$) part of the quark propagator. Both structures are only present when chiral 
symmetry is broken. If they are absent, as in the case of restored chiral symmetry, 
one obtains a different solution. We analyse this case below.

Before we discuss the solution (\ref{power}) further it is important to check that 
it persists to all orders in the skeleton expansion. Higher order terms in this 
expansion can be generated by inserting diagrammatical pieces like 
\beq
\epsfig{file=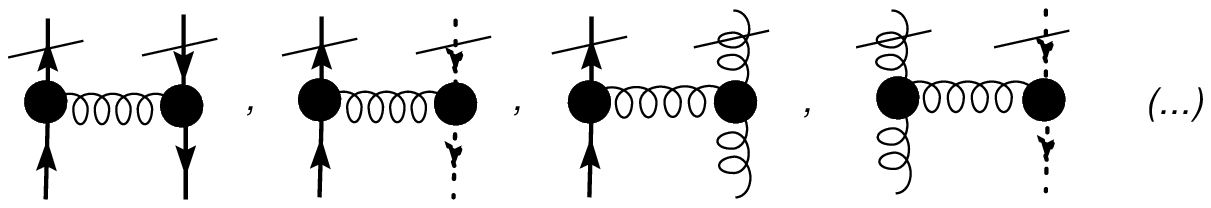,width=80mm} \nonumber
\eeq
into lower order diagrams (all propagators are fully dressed). In ref.\
\cite{Alkofer:2004it} we showed that insertions involving pure Yang-Mills propagators 
and vertices do not change the overall infrared divergence of a given diagram. This 
is also true for the three pieces involving quark lines: The first one introduces 
two quark-gluon vertices, two quark propagators and one gluon propagator plus a 
new integration into a given diagram. This amounts to an additional power of
\beq
\frac{Z_f^2 \,\, (p^2)^{-1-2\kappa + 2\kappa-1 + 2}}{M^2} = \frac{Z_f^2}{M^2}
\eeq
in the external momentum $(p^2)$. A similar result is obtained for the other pieces. 
Thus diagrammatic insertions of quark lines do not change the infrared behaviour of 
a given diagram but introduce additional powers of $Z_f/M$. This implies that the 
solution (\ref{power}) is valid to all orders in the skeleton expansion. It is 
therefore also an infrared solution of the full vertex-DSE. 

We wish to add that the infrared divergence presented here has been found under
the pretext that all external momenta of the vertex go to zero. However, a more 
detailed analysis of the vertex-DSE also shows, that the power law must be more 
general, $(p_3^2)^{-1/2-\kappa}$ in terms of the gluon momentum $p_3$. The uniform 
divergence presented here is therefore a particular case since, once all momenta 
vanish, also the gluon momentum vanishes. The details of the more general analysis 
will be presented elsewhere \cite{bigpaper}. 

\begin{figure}[b!]
\centerline{\epsfig{file=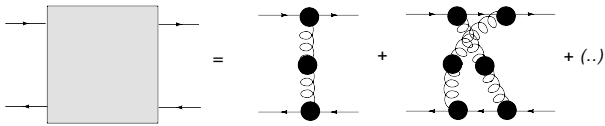,width=80mm}}
\caption{The four-quark 1PI Green's function and the first terms of its skeleton 
expansion}
\label{qq-1PI}
\end{figure}

An important application of our infrared power counting scheme concerns the 
four-quark 1PI Green's function $H(p)$, which is given in Fig.~\ref{qq-1PI} 
together with its skeleton expansion. From the terms of the skeleton
expansion one obtains $H(p) \sim (p^2)^{-2}$ in the infrared. 
The well known relation
\beq
V({\bf r}) = \frac{1}{(2\pi)^3} \int H(p^0=0,{\bf p}) e^{i {\bf p r}} d^3p 
\ \ \sim \ \ |{\bf r} | \label{potential}
\eeq
between the static four-quark function $H(p^0=0,{\bf p})$ and the quark 
potential $V({\bf r})$ therefore gives a linear rising potential by naive 
dimensional arguments. A more refined treatment, as described in 
\cite{Gromes:1981cb}, leads to the same result. Note that already the 
first term in the skeleton expansion, {\it i.e.\/} nonperturbative one-gluon 
exchange displayed in Fig.~\ref{qq-1PI}, generates this result. Since the 
following terms in the expansion are equally enhanced in the infrared, the 
string tension will be built up by summing over an infinite number of
diagrams. The coefficients of these diagrams will be investigated 
elsewhere \cite{bigpaper}.

A further interesting quantity is the running coupling $\alpha_{qg}$ from the 
quark-gluon vertex. A nonperturbative and renormalisation group invariant 
definition of this coupling is given by
\beq
\alpha_{qg}(p^2) = \frac{g^2}{4\pi} \, \xi_1^2(p^2) \, Z_f^2(p^2) \, 
Z(p^2) \, \sim \, \frac{1}{N_c} \frac{1}{p^2}\,,
\eeq
where $\xi_1$ dresses the $\gamma_\mu$-part of the vertex, see Eq.~(\ref{vertex}). 
With the exponents $\beta_2=\beta_1=-1/2-\kappa$, as obtained in our numerical 
solutions, we find by power counting that the coupling is proportional to $1/p^2$. 
Thus, contrary to the couplings (\ref{alpha}) from the Yang-Mills vertices, we 
find this coupling to be singular in the infrared,
i.e. 'infrared slavery' is realised.

So far we found quark confinement in the chirally broken phase
of quenched QCD. Our power counting scheme worked in this case, because only
one external momentum and no other small scales were present. This is also true in
the chirally symmetric phase of (quenched) QCD with massless  valence quarks.
The solution (\ref{IRsolution}) for the pure Yang-Mills sector is still valid 
in this case. In the quark sector all tensor structures in the quark propagator
and the quark-gluon vertex that violate chiral symmetry have to disappear, in
particular  $\xi_2$ and $\xi_4$ have to vanish identically. We are thus
left with
\beq
S(p) = \frac{i \pslash}{p^2}Z_f(p^2) \,, \ \ \ \Gamma_\mu (p) = 
i g \sum_{i=1,3} \xi_i (p^2) G_{\mu}^i 
\eeq
for the quark propagator and for the quark-gluon vertex. On general grounds 
$Z_f(p^2)$ cannot be infrared singular and can be treated as a constant. We 
then obtain 
\beq
\beta_1 = \beta_3 = - \kappa
\eeq
for the exponents in (\ref{vertex-exp}). This solution no longer leads to a 
confining potential. 
One obtains $H(p) \sim (p^2)^{-1}$ for the infrared behaviour of the
four-quark function. This leads to a potential of Coulomb type
\beq
V({\bf r}) = \frac{1}{(2\pi)^3} \int H(p^0=0,{\bf p}) e^{i {\bf p r}} d^3p 
\ \ \sim \ \ \frac{1}{|{\bf r} |}\,.
\eeq
Also the resulting running coupling from the quark-gluon vertex is no longer 
diverging but goes to a fixed point in the infrared 
\beq
\alpha_{qg}(p^2) = \frac{g^2}{4\pi} \, \xi_1^2(p^2) \, Z_f^2(p^2) \, 
Z(p^2) \, \sim \, \frac{1}{N_c} \,,
\eeq
similar to the couplings from the Yang-Mills vertices. The restoration of chiral 
symmetry is therefore directly linked with the disappearance of infrared 
slavery and confinement.

To summarise: We presented an analysis of quenched QCD in the covariant Landau
gauge. Our results show that infrared slavery is at work, though in a different
fashion than has  been assumed in many previous investigations. 
It is not a gluon propagator with
a $1/p^4$-behaviour in the infrared  that confines quarks. Instead the gluon
propagator is even vanishing at zero momentum. However,  there is enough
infrared strength in the quark-gluon vertex to compensate for this: We found a
linear rising potential from the four-quark Green's function. This
potential is triggered by scalar Dirac amplitudes in the quark propagator and
quark-gluon vertex. However, since  $\beta_1=\beta_2$, the potential also
contains vector contributions. An answer to the old  question of scalar vs.
vector confinement is therefore a nontrivial dynamical issue. Corresponding
numerical results will be published elsewhere \cite{bigpaper}.
Finally, we wish to emphasise again that if chiral symmetry is restored the 
confining solution disappears. 
As a result we have uncovered a novel link between dynamical chiral symmetry 
breaking and confinement.

\smallskip
{\bf Acknowledgements}\\
This work has been supported by the Deutsche For\-schungsgemeinschaft (DFG) 
under contracts Al 279/5-1 and Fi 970/7-1 and 
the Spanish grant FPA 2004-02602.

\end{document}